\documentclass{icrc29}
\usepackage{graphicx,amssymb,amsmath,times}
\begin{document}
%Title of paper
\title[Fluorescence in damp air  ...... ] 
{Fluorescence in damp air and comments on the radiative life time} 
\author[N.Sakaki et al.]{N. Sakaki$^a$, K. Kobayakawa$^b$,
       M. Nagano$^c$ and Y. Watanabe$^c$ \\
   (a) Department of Physics and Mathematics, Aoyama Gakuin University,
     Sagamihara, 229-8558 Japan \\
   (b) Department of of Architecture and Civil Engineering, Fukui University
   of Technology, Fukui, 910-8505 Japan \\
   (c) Department of Space Communication Engineering, Fukui University
   of Technology, Fukui, 910-8505 Japan \\
}
\presenter{Presenter: N. Sakaki (sakaki@phys.aoyama.ac.jp), \
jap-sakaki-N-abs1-he15-oral}

\maketitle

\begin{abstract}
Photon yields in damp air excited by an electron using a
$^{90}$Sr $\beta$ source are compared with
those in dry air.   
Water vapors considerably reduce the yields, however, a 
further study is needed to evaluate the effects on the energy
estimation of ultrahigh-energy cosmic rays.
The relation of fluorescence efficiency to the life time of
de-excitation by radiation is discussed.
\end{abstract}

\section{Introduction}
Photon yields in damp air are quite important to estimate
the primary energy of cosmic rays with the fluorescence
technique from space like EUSO. After our previous measurement
in dry air \cite{nag04,nag03}, the measurement has been
continued using  $^{90}$Sr $\beta$ source to study
the pressure dependence of photon yields for radiation
in damp air.  The results in 21\% and 56\% relative humidities
at one atmosphere are presented in section 3.

In our previous report \cite{nag03}, we were concerned with the decay time,
$\tau$. It consists of three terms: (1) the mean lifetime 
of the excited state to the lower state, $\tau_r$, with which the 
fluorescence radiates; (2) that of internal quenching (internal conversion 
plus inter-system crossing) , $\tau_q$ ; and (3)  the lifetime of 
collisional de-excitation, $\tau_c$ which is proportional to $\sqrt{T}/p$,
where $T$ and $p$ are the temperature and the pressure, respectively.
We have used the fluorescence efficiency; i.e.
$\mathit{\Phi}(p) = \tau /\tau_r $, and
$\mathit{\Phi}^{\circ} = \tau_o /\tau_r $ , where $\tau_o $ is 
the combined lifetime of $\tau_r$ and $\tau_q $ .

Since  $\mathit{\Phi}^{\circ}$ was in order of 10$^{-3}$ and $\tau_o$ 
was in order of a few tens ns, these led that $\tau_r$ would be in order of
a few ten $\mu$s. This value of $\tau_r$ is quite large compared with ones
obtained so far. 
Thus we will correct our relation between $\mathit{\Phi} ^{\circ}$ and the
photon yield $\epsilon$ in the next section.  We will discuss them in more 
detail in section {\bf 4}. 

\section{Photon yield}

The photon yield per unit length per electron for the $i$th band 
at pressure $p$, $\epsilon _i (p)$ can be written by
\begin{equation}
\epsilon _i (p) = \rho \frac{\mathrm{d}E}{\mathrm{d}x} \left( \frac{1}
{h\nu_i} \right) \cdot \varphi_i (p) \quad, \mbox{where} \ \ 
\varphi_i (p) = \kappa_i \mathit{\Phi}_i(p).
%\label{eq-eps}
\end{equation}
In eq.(1) $\mathit{\Phi}_i(p)$ is the fluorescence efficiency and
we call $\varphi_i(p)$ the modified efficiency. The inverse
of $\varphi_i (p)$ is expressed by
\begin{equation}
\frac{1}{\varphi_i(p)}=\frac{1}{\kappa_i \mathit{\Phi}_i^{\circ}}
\left(1+ \frac{p}{p'_i} \right),
%\label{eq-eff}
\end{equation}

$p'_i$ is the reference pressure where $\tau_{ci}$ is equal to 
$\tau_{oi}$ , namely $1/\tau_{ci}=(1/\tau_{oi})(p/p'_i)$.
 $ \mathit{\Phi}_i^{\circ} = \tau_{oi}/\tau_{ri}$ as before.
 $\kappa_i$ in eq.(1) is introduced here different from
refs \cite{nag04,nag03,elb93}. Its physical meaning will be explained 
further in section {\bf 5}. The eq.(1) can be written as a function
of pressure as
\begin{equation}
\epsilon _i (p) = \frac{Cp}{1+ \frac{p}{p'_i}} \quad . 
\end{equation}

\section{Experiment}
The experimental details are described in \cite{nag04,nag03}.
The central values of the narrow band filters used in the present measurement
are 337.7, 356.3 and 392.0~nm and their bandwidths at half maximum
are 9.8, 9.3 and 4.35~nm, respectively.   
The measurements have been done 
in dry air (mixing of 79\% nitrogen and 21\% oxygen) and
in normal air with (21$\pm$4)\% and (56$\pm$5)\% 
relative humidities (RH) at one atmosphere.
The pressure was varied from one atmosphere to 10 hPa under
the temperature of (19.7$\pm0.5)^{\circ}$C, keeping the specific
humidity (SH : ratio of mass of the vapor to that of damp air) 
to be constant.
The photon yield  $\epsilon$ is determined from the number of
signal counts, the total number of electrons, the length of the
fluorescence portion, the solid angle of the photomultiplier(PMT),
the quartz window transmission, the filter transmission and the
quantum efficiency and the collection efficiency of the PMT. 

\begin{figure}[thp]
\begin{center}
\includegraphics*[width=0.70\textwidth,angle=0,clip]{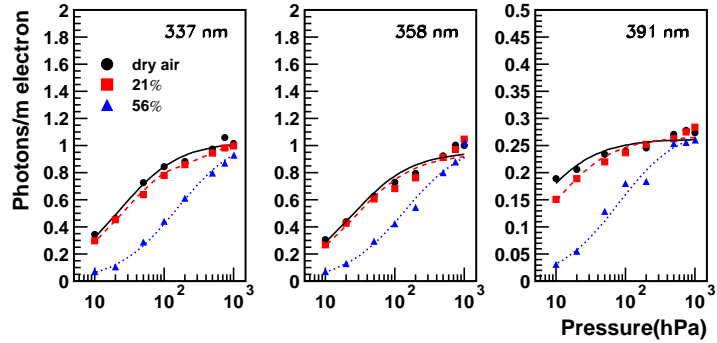}
\caption{\label {fig-yield}
The pressure dependence of $\epsilon$ in dry air and
damp air with RH, 21\% and 56\% at one atmosphere.
The temperature is kept constant (19.7$\pm0.5)^{\circ}$C throughout
the measurements. In each series of measurements, SH
(0, 3.0 and 8.0 g/kg, respectively) is kept constant, though RH changes. 
}
\end{center}
\end{figure}

The pressure dependence of $\epsilon$ in dry and damp air 
is shown in Figure \ref{fig-yield} for three
filter bands (We designate each filter band as  337, 358 and 391~nm).  
In these plots, the contribution of
different bands in one filter band width are not separated,
which is different from the results in our previous
paper \cite{nag04}. 
The results of
least square(LS) fitting to the eq.(3) are shown by solid 
curves (dry), dashed curves (21\%) and
dotted curves (56\%).

\begin{figure}[thp]
\begin{center}
\includegraphics*[width=0.70\textwidth,angle=0,clip]{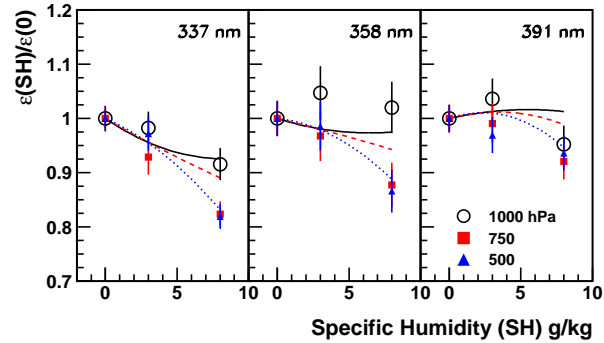}
\caption{\label {fig-ratio}
The ratio of $\epsilon$ at SH=0, 3.0 and 8.0 g/kg
to that of  $\epsilon$ of dry air(SH=0) are plotted as a function of SH
at three pressures.
}
\end{center}
\end{figure}

The ratio of $\epsilon$ at SH=0, (3.0$\pm$0.5) and (8.0$\pm$0.7) g/kg
to that of  $\epsilon$ of dry air(SH=0) are plotted as a function of SH
at 1000, 750 and 500 hPa in Figure \ref{fig-ratio}.
Calculated ratios with the LS fitted $p'$ and $C$ values
determined above are shown by a solid, dashed and dotted curves for
1000, 750 and 500 hPa.
The SH dependencies for 337~nm and 358~nm (2P transition) seem to be 
similar, but different from that for 391~nm (1N transition). 
We are repeating and continuing the measurements with different RHs
to make the effect of water vapor more clear.

\begin{figure}[thp]
\begin{center}
\includegraphics*[width=0.70\textwidth,angle=0,clip]{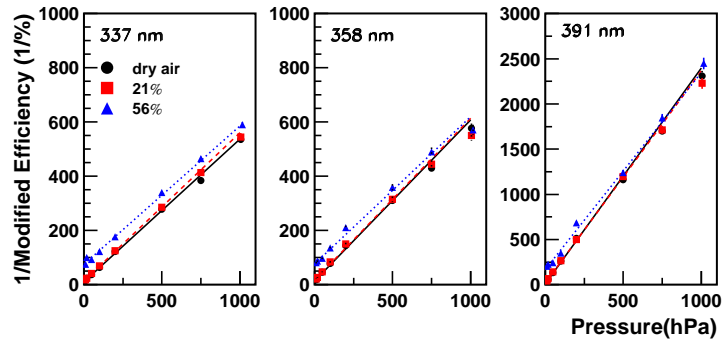}
\caption{\label {fig-eff}
The reciprocal of the modified efficiency 
$\frac{1}{\varphi_i(p)}$ in dry 
and damp air are plotted as a function of pressure. 
The LS-fit
lines are also shown.
}
\end{center}
\end{figure}

The reciprocal of the modified efficiency 
$\frac{1}{\varphi_i(p)}$
 for air in $\frac{1}{\%}$ are plotted 
as a function of pressure in Figure \ref{fig-eff}.  The results of
LS fitting to eq.(2) are shown by solid (dry), 
dashed (21\%) and dotted (56\%) lines.

In the present measurement the temperature was kept constant
for decreasing the pressure.  Actually, the atmospheric temperature 
decreases with the lapse-rate of about 6.5 K/km.
The application of the present
result to the real atmosphere is under study. 

\section{Notes on radiative life time and photon yield of fluorescence}
In eq.(1), a new parameter $\kappa_i$ is introduced. If $\kappa_i =1$ ,
then $\varphi_i (p) = \mathit{\Phi}_i(p) $. In this case, eq.(1) holds 
only under very simple assumptions, namely (a) a charged particle is lost 
its energy ${\rm d}E$ by only one process, i.e. exciting 
an electron in molecule from the ground state to the only one excited state. 
And the excited electron emits $h\nu_i$ light or internally converts: 
${\rm d}E/h\nu_i$  represents the number of photons. Of course, this is not 
the case in air as well as ${\rm N}_2$ and  ${\rm N}_2^+$. 

The 2P transition does not take two steps, i.e. excitation and 
radiation. In 2P, the direct process from the ground state to the
specific electronic state
${\rm C}\,^3{\rm \Pi}_u$ is forbidden because of the change of spin 
multiplet. There are many excited levels in ${\rm N}_2$. And 2P 
radiative transition is not to the ground state. 
So the exciting energy is 11.03 eV while the radiation 
energy $h\nu_i$ is 3.68 eV for, say, 2P(0,0). While in 1N, nitrogen 
gas must be, of course, ionized at first. 
There are also many other excited electronic states of 
${\rm N}_2^+$ than ${\rm B} \,^2 {\rm \Sigma}_u$.
 These facts lead that the 
simple assumptions mentioned above are not appropriate.
The energy loss ${\rm d}E$ of charged particles may mainly be caused 
by ionization rather than excitation. 

By referring Chapter 14 in \cite{bir64}, the factor $\kappa_i$ in 
eq.(1) is inserted. This factor means the rate of the total energy 
loss ${\rm d}E$  with which an electron in a 
molecule is excited to the specific electronic state with a definite 
vibration state, such as  ${\rm C}\,^3{\rm \Pi}_u$ 
electronic state with $v' = 1$ even indirectly. $\kappa_i$ also
contains a coefficient of proportion which may include the branching 
ratio of $B^{v',v''}$ \cite{gil92}, where $v'$ and $v''$ are initial 
and final vibration levels, respectively, and the energy ratio of 
radiation to excitation and so on. 

 So, values of $\kappa_i$ is considered to be much less than unity, say
$10^{-3}$.  
The values of $\mathit{\Phi}_i^{\circ}$ in eq.(2) 
are in order of unity. Then the values of $\tau_r$ is the order of a 
few tens ns which is consistent with the old but more accurate $\tau_r$ 
for 2P \cite{ben78,dot73} and for 1N \cite{dot73} than ours 
\cite{nag03}.  
It is noted that the comments described here on $\tau_0$ and $\tau_r$ 
do not have any effect on the analysis of the photon yield, $\epsilon$.

\section{Conclusion}
Photon yields in damp air under the constant temperature 
19.7$^{\circ}$C were investigated and compared with those in dry air.
Though the rate of the mole number of water to the total mole number
is small in damp air, water vapors seem to take a role in reducing 
the photon yields. We will continue to measure yields in damp air
under various conditions.

\bigskip
We are grateful to Joon C. Lee of University of Southern Mississippi
for raising the question on $\mathit{\Phi}^{\circ}$ 
and Fernando Arqueros of Universidad Complutense de Madrid 
for useful comments and discussions.

\end{document}